\documentclass{article}

\usepackage{arxiv}

\usepackage[utf8]{inputenc} 
\usepackage[T1]{fontenc}    
\usepackage{hyperref}       
\usepackage{url}            
\usepackage{booktabs}       
\usepackage{amsfonts}       
\usepackage{nicefrac}       
\usepackage{microtype}      
\usepackage{lipsum}		
\usepackage{graphicx}
\usepackage[semicolon]{natbib}
\usepackage{doi}
\usepackage{color}
\usepackage{fancyvrb}

\DefineVerbatimEnvironment{Highlighting}{Verbatim}{commandchars=\\\{\}}
\usepackage{framed}
\definecolor{shadecolor}{RGB}{248,248,248}
\newenvironment{Shaded}{\begin{snugshade}}{\end{snugshade}}

\newcommand{\AttributeTok}[1]{\textcolor[rgb]{0.13,0.29,0.53}{#1}}

\newcommand{\CommentTok}[1]{\textcolor[rgb]{0.56,0.35,0.01}{\textit{#1}}}

\newcommand{\ConstantTok}[1]{\textcolor[rgb]{0.56,0.35,0.01}{#1}}
\newcommand{\ControlFlowTok}[1]{\textcolor[rgb]{0.13,0.29,0.53}{\textbf{#1}}}

\newcommand{\DecValTok}[1]{\textcolor[rgb]{0.00,0.00,0.81}{#1}}

\newcommand{\FloatTok}[1]{\textcolor[rgb]{0.00,0.00,0.81}{#1}}
\newcommand{\FunctionTok}[1]{\textcolor[rgb]{0.13,0.29,0.53}{\textbf{#1}}}

\newcommand{\NormalTok}[1]{#1}

\newcommand{\OtherTok}[1]{\textcolor[rgb]{0.56,0.35,0.01}{#1}}

\newcommand{\SpecialCharTok}[1]{\textcolor[rgb]{0.81,0.36,0.00}{\textbf{#1}}}

\newcommand{\StringTok}[1]{\textcolor[rgb]{0.31,0.60,0.02}{#1}}

\makeatletter
 \let\@cite@ofmt\@firstofone
 \def\@biblabel#1{}
 \def\@cite#1#2{{#1\if@tempswa , #2\fi}}
\makeatother
\newlength{\cslhangindent}
\newlength{\csllabelwidth}
 {\begin{list}{}{%
  \setlength{\itemindent}{0pt}
  \setlength{\leftmargin}{0pt}
  \setlength{\parsep}{0pt}
  \ifodd #1
   \setlength{\leftmargin}{\cslhangindent}
   \setlength{\itemindent}{-1\cslhangindent}
  \fi
  \setlength{\itemsep}{#2\baselineskip}}}
 {\end{list}}
\usepackage{calc}

\title{fairmetrics: An R package for group fairness evaluation}

\author{ \href{https://orcid.org/0009-0007-2206-0177}{\includegraphics[scale=0.06]{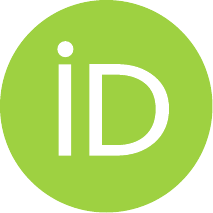}\hspace{1mm}Benjamin~Smith} \\
	Department of Statistical Sciences\\
	University of Toronto\\
	Toronto, ON M5G 1X6 \\
	\texttt{benyamindsmith@mail.utoronto.ca} \\
	\And
	\href{https://orcid.org/0000-0003-0915-1473}{\includegraphics[scale=0.06]{orcid.pdf}\hspace{1mm}Jianhui~Gao} \\
	Department of Statistical Sciences\\
	University of Toronto\\
	Toronto, ON M5G 1X6 \\
	\texttt{jianhui.gao@mail.utoronto.ca} \\
        \And
	\href{https://orcid.org/0000-0002-5360-5869}{\includegraphics[scale=0.06]{orcid.pdf}\hspace{1mm}Jessica~Gronsbell} \\
	Department of Statistical Sciences\\
	University of Toronto\\
	Toronto, ON M5G 1X6 \\
	\texttt{j.gronsbell@utoronto.ca} \\
}

\renewcommand{\shorttitle}{fairmetrics: An R package for group fairness evaluation}

\hypersetup{
pdftitle={fairmetrics: An R package for group fairness evaluation},
pdfsubject={stat.ML},
pdfauthor={Benjamin Smith, Jianhui Gao, Jessica Gronsbell},
pdfkeywords={Fairness, Machine Learning, Predictive Models, R Package},
}

\begin{document}
\maketitle

\section{Summary}\label{summary}

Fairness is a growing area of machine learning (ML) that focuses on
ensuring models do not produce systematically biased outcomes for
specific groups, particularly those defined by protected attributes such
as race, gender, or age. Evaluating fairness is a critical aspect of ML
model development, as biased models can perpetuate structural
inequalities. The \{fairmetrics\} R package offers a user-friendly
framework for rigorously evaluating numerous group-based fairness
criteria, including metrics based on independence (e.g., statistical
parity), separation (e.g., equalized odds), and sufficiency (e.g.,
predictive parity). Group-based fairness criteria assess whether a model
is equally accurate or well-calibrated across a set of predefined groups
so that appropriate bias mitigation strategies can be implemented.
\{fairmetrics\} provides both point and interval estimates for multiple
metrics through a convenient wrapper function and includes an example
dataset derived from the Medical Information Mart for Intensive Care,
version II (MIMIC-II) database \citep{goldberger2000physiobank, raffa2016clinical}.

\section{Statement of Need}\label{statement-of-need}

ML models are increasingly integrated into high-stakes domains to
support decision making that significantly impacts individuals and
society more broadly, including criminal justice, healthcare, finance,
employment, and education \citep{mehrabi_survey_21}. Mounting evidence
suggest that these models often exhibit bias across groups defined by
protected attributes. For example, within criminal justice, the
Correctional Offender Management Profiling for Alternative Sanctions
(COMPAS) software, a tool used by U.S. courts to evaluate the risk of
defendants becoming recidivists, was found to incorrectly classify Black
defendants as high-risk at nearly twice the rate of white defendants
\citep{mattuMachineBias}. This bias impacted Black defendants by potentially leading
to harsher bail decisions, longer sentences, and reduced parole
opportunities compared to white defendants with similar risk profiles.
Similarly, within healthcare, a commercial risk-prediction algorithm
deployed in the U.S. to identify patients with complex health needs for
high-risk care management programs was shown to be significantly less
calibrated for Black patients relative to white patients \citep{obermeyerDissectingRacialBias2019}. This caused Black patients with equivalent health conditions
to be under-referred for essential care services compared to white
patients. These examples illustrate that there is an urgent need for
practitioners and researchers to ensure that ML models support fair
decision making before they are deployed in real-world applications.

While existing software can compute group fairness criteria, they only
provide point estimates and/or visualizations without quantifying the
uncertainty around the criteria. This limitation prevents users from
determining whether observed disparities between groups are
statistically significant or merely the result of random variation due
to finite sample size, potentially leading to incorrect conclusions
about fairness violations. The \{fairmetrics\} R package addresses this
gap by providing bootstrap-based confidence intervals (CIs) for both
difference-based and ratio-based group fairness metrics, empowering
users to make statistically grounded decisions about the fairness of
their models, which is inconsistently done in practice.

\section{Scope}\label{scope}

The \texttt{\{fairmetrics\}} package is designed to evaluate group
fairness in the setting of binary classification with a binary protected
attribute. This restriction reflects standard practice in the fairness
literature and is motivated by several considerations. First, binary
classification remains prevalent in many high-stakes applications, such
as loan approval, hiring decisions, and disease screening, where
outcomes are typically framed as accept/reject or positive/negative
\citep{mehrabi_survey_21}. Second, group fairness is the most widely used
framework for binary classification tasks \citep{mehrabi_survey_21}. Third,
when protected attributes have more than two categories, there is no
clear consensus on how to evaluate group fairness \citep{lum_debias_22}. This focus enables \{fairmetrics\} to provide statistically
grounded uncertainty quantification for group fairness metrics commonly
applied in binary classification tasks across diverse application
domains.

\section{Fairness Criteria}\label{fairness-criteria}

Group fairness criteria are primarily classified into three main
categories: independence, separation, and sufficiency \citep{barocas2023fairness, Berk_Heidari_Jabbari_Kearns_Roth_2018, Castelnovo_Crupi_Greco_Regoli_Penco_Cosentini_2022, Gao_Chou_McCaw_Thurston_Varghese_Hong_Gronsbell_2024}. Independence requires that the model's classifications be
statistically independent of the protected attribute, meaning the
likelihood of receiving a positive prediction is the same across
protected groups. Separation requires independence between the
classifications and the protected attribute conditional on the true
outcome, so that the probability of a positive prediction is equal
across protected groups within the positive (or negative) outcome class.
Sufficiency requires independence between the outcome and the protected
attribute conditional on the prediction, implying that once the model's
prediction is known, the protected attribute provides no additional
information about the true outcome. Below we summarize the fairness
metrics that are available within the \{fairmetrics\} package.

\subsection{Independence}\label{independence}

\begin{itemize}
\item
  \textbf{Statistical Parity:} Compares the overall rate of positive
  predictions between groups.
\end{itemize}

\subsection{Separation}\label{separation}

\begin{itemize}
\item
  \textbf{Equal Opportunity:} Compares disparities in the false negative
  rates between groups, quantifying differences in the likelihood of
  missing positive outcomes.
\item
  \textbf{Predictive Equality:} Compares the false positive rates (FPR)
  between groups, quantifying differences in the likelihood of
  incorrectly labeling negative outcomes as positive.
\item
  \textbf{Balance for Positive Class:} Compares the average of the
  predicted probabilities among individuals whose true outcome is
  positive across groups.
\item
  \textbf{Balance for Negative Class:} Compares the average of the
  predicted probabilities among individuals whose true outcome is
  negative across groups.
\end{itemize}

\subsection{Sufficiency}\label{sufficiency}

\begin{itemize}
\item
  \textbf{Positive Predictive Parity:} Compares the positive predictive
  values across groups, assessing differences in the precision of
  positive predictions.
\item
  \textbf{Negative Predictive Parity:} Compares the negative predictive
  values across groups, assessing differences in the precision of
  negative predictions.
\end{itemize}

\subsection{Other Criteria}\label{other-criteria}

\begin{itemize}
\item
  \textbf{Brier Score Parity:} Compares the Brier score (i.e., the mean
  squared error of the predicted probabilities) across groups,
  evaluating differences in calibration.
\item
  \textbf{Accuracy Parity:} Compares the overall accuracy of a
  predictive model across groups.
\item
  \textbf{Treatment Equality:} Compares the ratio of false negatives to
  false positives across groups, evaluating whether the trade-off
  between missed detections of positive outcomes and false alarms of
  negative outcomes is balanced.
\end{itemize}

\section{Evaluating Fairness
Criteria}\label{evaluating-fairness-criteria}

\begin{figure}
\centering
\includegraphics[width=\textwidth]{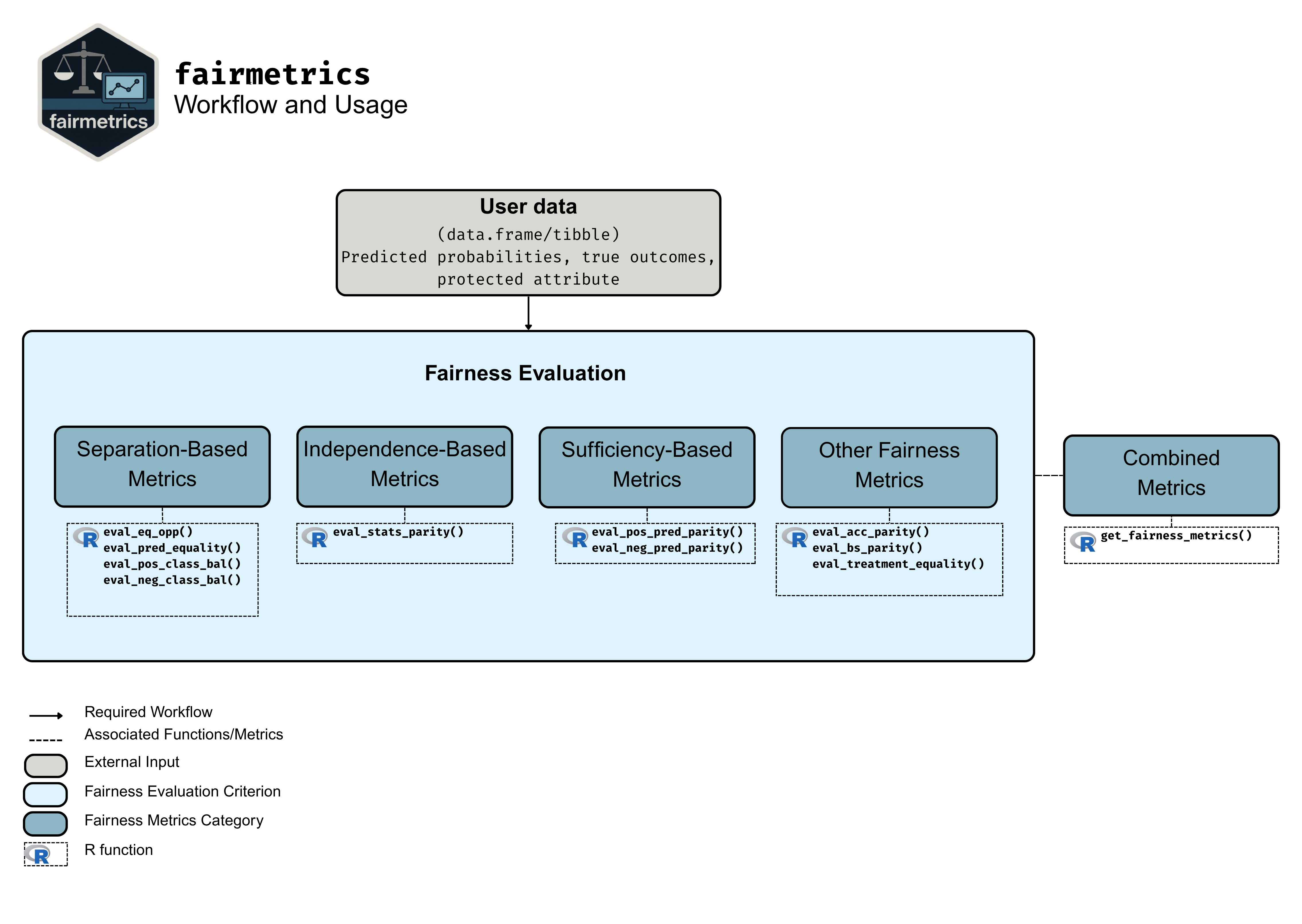}
\caption{Workflow for using \{fairmetrics\} to evaluate model fairness
across multiple criteria. \label{workflow}}
\end{figure}

The input to the \{fairmetrics\} package is a data frame or tibble
containing the model's predicted probabilities, the true outcomes, and
the protected attribute of interest.
\hyperref[workflow]{Figure ~\ref*{workflow}} shows the workflow for
using \{fairmetrics\}. Users can evaluate a model for a specific
criterion or multiple group fairness criteria using the combined metrics
function.

A simple example of how to use the \{fairmetrics\} package is
illustrated below. The example makes use of the
\texttt{mimic\_preprocessed} dataset, a pre-processed version of the the
Indwelling Arterial Catheter (IAC) Clinical Dataset, from the MIMIC-II
clinical database\footnote{The raw version of this data is made
  available by PhysioNet \citep{goldberger2000physiobank} and can be accessed in
  the \{fairmetrics\} package by loading the \texttt{mimic} dataset.}
\citep{raffa2016clinical, raffa2016data}. The dataset consists of 1,776
hemodynamically stable patients with respiratory failure and includes
demographic information (patient age and gender), vital signs,
laboratory results, whether an IAC was used, and a binary outcome
indicating whether the patient died within 28 days of hospital
admission.

While the choice of fairness criteria used is context dependent, we show
all criteria available with the \texttt{get\_fairness\_metrics()}
function for the purposes of illustration. In this example, we evaluate
the model's fairness with respect to the protected attribute
\texttt{gender}. For conditional statistical parity, we condition on
patients older than 60 years old. The model is trained on a subset of
the data and the predictions are made and evaluated on a test set. The
\texttt{get\_fairness\_metrics()} function outputs difference and
ratio-based metrics as well as their corresponding confidence intervals.
A statistically significant difference across groups at a given level of
significance is indicated when the confidence interval for a
difference-based metric does not include zero or when the interval for a
ratio-based metric does not include one.

\begin{Shaded}
\begin{Highlighting}[]
\FunctionTok{library}\NormalTok{(fairmetrics)}
\FunctionTok{library}\NormalTok{(dplyr)}
\FunctionTok{library}\NormalTok{(magrittr)}
\FunctionTok{library}\NormalTok{(randomForest)}

\CommentTok{\# Load the example dataset}
\FunctionTok{data}\NormalTok{(}\StringTok{"mimic\_preprocessed"}\NormalTok{)  }

\CommentTok{\# Split the data into training and test sets}
\NormalTok{train\_data }\OtherTok{\textless{}{-}}\NormalTok{ mimic\_preprocessed }\SpecialCharTok{\%\textgreater{}\%}
\NormalTok{  dplyr}\SpecialCharTok{::}\FunctionTok{filter}\NormalTok{(dplyr}\SpecialCharTok{::}\FunctionTok{row\_number}\NormalTok{() }\SpecialCharTok{\textless{}=} \DecValTok{700}\NormalTok{)}

\NormalTok{test\_data }\OtherTok{\textless{}{-}}\NormalTok{ mimic\_preprocessed }\SpecialCharTok{\%\textgreater{}\%}
\NormalTok{  dplyr}\SpecialCharTok{::}\FunctionTok{mutate}\NormalTok{(}\AttributeTok{gender =} \FunctionTok{ifelse}\NormalTok{(gender\_num }\SpecialCharTok{==} \DecValTok{1}\NormalTok{, }\StringTok{"Male"}\NormalTok{, }\StringTok{"Female"}\NormalTok{)) }\SpecialCharTok{\%\textgreater{}\%}
\NormalTok{  dplyr}\SpecialCharTok{::}\FunctionTok{filter}\NormalTok{(dplyr}\SpecialCharTok{::}\FunctionTok{row\_number}\NormalTok{() }\SpecialCharTok{\textgreater{}} \DecValTok{700}\NormalTok{)}

\CommentTok{\# Train a random forest model}
\NormalTok{rf\_model }\OtherTok{\textless{}{-}}\NormalTok{ randomForest}\SpecialCharTok{::}\FunctionTok{randomForest}\NormalTok{(}
  \FunctionTok{factor}\NormalTok{(day\_28\_flg) }\SpecialCharTok{\textasciitilde{}}\NormalTok{ ., }
  \AttributeTok{data =}\NormalTok{ train\_data, }
  \AttributeTok{ntree =} \DecValTok{1000}
\NormalTok{  )}
  
\CommentTok{\# Make predictions on the test set}
\NormalTok{test\_data}\SpecialCharTok{$}\NormalTok{pred }\OtherTok{\textless{}{-}} \FunctionTok{predict}\NormalTok{(rf\_model, }\AttributeTok{newdata =}\NormalTok{ test\_data, }\AttributeTok{type =} \StringTok{"prob"}\NormalTok{)}

\CommentTok{\# Get fairness metrics}
\CommentTok{\# Setting alpha=0.05 for 95\% confidence intervals}
\FunctionTok{get\_fairness\_metrics}\NormalTok{(}
 \AttributeTok{data =}\NormalTok{ test\_data,}
 \AttributeTok{outcome =} \StringTok{"day\_28\_flg"}\NormalTok{,}
 \AttributeTok{group =} \StringTok{"gender"}\NormalTok{,}
 \AttributeTok{group2 =} \StringTok{"age"}\NormalTok{,}
 \AttributeTok{probs =} \StringTok{"pred"}\NormalTok{,}
 \AttributeTok{cutoff =} \FloatTok{0.41}\NormalTok{, }
 \AttributeTok{alpha =} \FloatTok{0.05}
\NormalTok{)}


\NormalTok{           Fairness Assesment                                  Metric}
\DecValTok{1}\NormalTok{          Statistical Parity                Positive Prediction Rate}
\DecValTok{2}\NormalTok{           Equal Opportunity                     False Negative Rate}
\DecValTok{3}\NormalTok{         Predictive Equality                     False Positive Rate}
\DecValTok{4}\NormalTok{  Balance }\ControlFlowTok{for}\NormalTok{ Positive Class           Avg. Predicted Positive Prob.}
\DecValTok{5}\NormalTok{  Balance }\ControlFlowTok{for}\NormalTok{ Negative Class           Avg. Predicted Negative Prob.}
\DecValTok{6}\NormalTok{  Positive Predictive Parity               Positive Predictive Value}
\DecValTok{7}\NormalTok{  Negative Predictive Parity               Negative Predictive Value}
\DecValTok{8}\NormalTok{          Brier Score Parity                             Brier Score}
\DecValTok{9}\NormalTok{     Overall Accuracy Parity                                Accuracy}
\DecValTok{10}\NormalTok{         Treatment }\FunctionTok{Equality}\NormalTok{ (False Negative)}\SpecialCharTok{/}\NormalTok{(False Positive) Ratio}

\NormalTok{   GroupFemale GroupMale Difference    }\DecValTok{95}\SpecialCharTok{\% Diff CI Ratio 95\%}\NormalTok{ Ratio CI}
\DecValTok{1}         \FloatTok{0.17}      \FloatTok{0.08}       \FloatTok{0.09}\NormalTok{   [}\FloatTok{0.05}\NormalTok{, }\FloatTok{0.13}\NormalTok{]  }\FloatTok{2.12}\NormalTok{ [}\FloatTok{1.49}\NormalTok{, }\FloatTok{3.04}\NormalTok{]}
\DecValTok{2}         \FloatTok{0.38}      \FloatTok{0.62}      \SpecialCharTok{{-}}\FloatTok{0.24}\NormalTok{ [}\SpecialCharTok{{-}}\FloatTok{0.39}\NormalTok{, }\SpecialCharTok{{-}}\FloatTok{0.09}\NormalTok{]  }\FloatTok{0.61}\NormalTok{ [}\FloatTok{0.44}\NormalTok{, }\FloatTok{0.86}\NormalTok{]}
\DecValTok{3}         \FloatTok{0.08}      \FloatTok{0.03}       \FloatTok{0.05}\NormalTok{   [}\FloatTok{0.02}\NormalTok{, }\FloatTok{0.08}\NormalTok{]  }\FloatTok{2.67}\NormalTok{  [}\FloatTok{1.4}\NormalTok{, }\FloatTok{5.08}\NormalTok{]}
\DecValTok{4}         \FloatTok{0.46}      \FloatTok{0.37}       \FloatTok{0.09}\NormalTok{   [}\FloatTok{0.04}\NormalTok{, }\FloatTok{0.14}\NormalTok{]  }\FloatTok{1.24}\NormalTok{ [}\FloatTok{1.09}\NormalTok{, }\FloatTok{1.42}\NormalTok{]}
\DecValTok{5}         \FloatTok{0.15}      \FloatTok{0.10}       \FloatTok{0.05}\NormalTok{   [}\FloatTok{0.03}\NormalTok{, }\FloatTok{0.07}\NormalTok{]  }\FloatTok{1.50}\NormalTok{ [}\FloatTok{1.29}\NormalTok{, }\FloatTok{1.74}\NormalTok{]}
\DecValTok{6}         \FloatTok{0.62}      \FloatTok{0.66}      \SpecialCharTok{{-}}\FloatTok{0.04}\NormalTok{  [}\SpecialCharTok{{-}}\FloatTok{0.21}\NormalTok{, }\FloatTok{0.13}\NormalTok{]  }\FloatTok{0.94}\NormalTok{ [}\FloatTok{0.72}\NormalTok{, }\FloatTok{1.22}\NormalTok{]}
\DecValTok{7}         \FloatTok{0.92}      \FloatTok{0.90}       \FloatTok{0.02}\NormalTok{  [}\SpecialCharTok{{-}}\FloatTok{0.02}\NormalTok{, }\FloatTok{0.06}\NormalTok{]  }\FloatTok{1.02}\NormalTok{ [}\FloatTok{0.98}\NormalTok{, }\FloatTok{1.07}\NormalTok{]}
\DecValTok{8}         \FloatTok{0.09}      \FloatTok{0.08}       \FloatTok{0.01}\NormalTok{  [}\SpecialCharTok{{-}}\FloatTok{0.01}\NormalTok{, }\FloatTok{0.03}\NormalTok{]  }\FloatTok{1.12}\NormalTok{ [}\FloatTok{0.89}\NormalTok{, }\FloatTok{1.43}\NormalTok{]}
\DecValTok{9}         \FloatTok{0.87}      \FloatTok{0.88}      \SpecialCharTok{{-}}\FloatTok{0.01}\NormalTok{  [}\SpecialCharTok{{-}}\FloatTok{0.05}\NormalTok{, }\FloatTok{0.03}\NormalTok{]  }\FloatTok{0.99}\NormalTok{ [}\FloatTok{0.94}\NormalTok{, }\FloatTok{1.04}\NormalTok{]}
\DecValTok{10}        \FloatTok{1.03}      \FloatTok{3.24}      \SpecialCharTok{{-}}\FloatTok{2.21}\NormalTok{ [}\SpecialCharTok{{-}}\FloatTok{4.38}\NormalTok{, }\SpecialCharTok{{-}}\FloatTok{0.04}\NormalTok{]  }\FloatTok{0.32}\NormalTok{ [}\FloatTok{0.15}\NormalTok{, }\FloatTok{0.68}\NormalTok{]}
\end{Highlighting}
\end{Shaded}

Should the user wish to calculate an individual criteria, it is possible
to use any of the \texttt{eval\_*} functions. For example, to calculate
equal opportunity, the user can call the
\texttt{eval\_equal\_opportunity()} function.

\begin{Shaded}
\begin{Highlighting}[]
\FunctionTok{eval\_eq\_opp}\NormalTok{(}
  \AttributeTok{data =}\NormalTok{ test\_data,}
  \AttributeTok{outcome =} \StringTok{"day\_28\_flg"}\NormalTok{,}
  \AttributeTok{group =} \StringTok{"gender"}\NormalTok{,}
  \AttributeTok{probs =} \StringTok{"pred"}\NormalTok{,}
  \AttributeTok{confint =} \ConstantTok{TRUE}\NormalTok{,}
  \AttributeTok{cutoff =} \FloatTok{0.41}\NormalTok{,}
  \AttributeTok{alpha =} \FloatTok{0.05}
\NormalTok{)}


\NormalTok{There is evidence that model does not satisfy equal opportunity.}
\NormalTok{               Metric GroupFemale GroupMale Difference    }\DecValTok{95}\SpecialCharTok{\% Diff CI Ratio 95\%}\NormalTok{ Ratio CI}
\DecValTok{1}\NormalTok{ False Negative Rate       }\SpecialCharTok{{-}}\FloatTok{0.42}     \SpecialCharTok{{-}}\FloatTok{0.23}      \SpecialCharTok{{-}}\FloatTok{0.19}\NormalTok{ [}\SpecialCharTok{{-}}\FloatTok{0.33}\NormalTok{, }\SpecialCharTok{{-}}\FloatTok{0.05}\NormalTok{]  }\FloatTok{1.83}\NormalTok{    [}\FloatTok{1.11}\NormalTok{, }\DecValTok{3}\NormalTok{]}
\end{Highlighting}
\end{Shaded}

\section{Related Work}\label{related-work}

Other R packages similar to \{fairmetrics\} include \{fairness\}
\citep{fairness_package}, \{fairmodels\} \citep{wisniewski2022fairmodels}
and \{mlr3fairness\} \citep{mlr3fairness_package}. The differences
between \{fairmetrics\} and these other packages is twofold. The primary
difference between is that \{fairmetrics\} calculates the ratio and
difference between group fairness criterion and their corresponding
confidence intervals of fairness metrics via bootstrap, allowing for
more meaningful inferences about the fairness criteria. Additionally, in
contrast to the \{fairmodels\}, \{fairness\} and \{mlr3fairness\}
packages, the \{fairmetrics\} package does not posses any external
dependencies and has a lower memory footprint, resulting in an
environment agnostic tool that can be used with modest hardware and
older systems.
\hyperref[tab:memory_dep_usage]{Table~\ref*{tab:memory_dep_usage}} shows
the comparison of memory used and dependencies required when loading
each library.

\begin{table}[ht]
\centering
\begin{tabular}{l r r}
\hline
\textbf{Package} & \textbf{Memory (MB)} & \textbf{Dependencies} \\
\hline
fairmodels  & 17.02  & 29 \\
fairness    & 117.61 & 141\\
mlr3fairness  & 58.11  & 45 \\
fairmetrics & 0.05   & 0  \\
\hline
\end{tabular}
\caption{Memory usage (in MB) and dependencies of {fairmetrics} vs similar packages.}
\label{tab:memory_dep_usage}
\end{table}

For python users, the \{fairlearn\} library \citep{fairlearn_paper}
provides additional fairness metrics and algorithms. The \{fairmetrics\}
package is designed for seemless integration with R workflows, making it
a more convenient choice for R-based ML applications.

\section{Licensing and Availability}\label{licensing-and-availability}

The \{fairmetrics\} package is under the MIT license. It is available on
CRAN and can be installed by using
\texttt{install.packages("fairmetrics")}. A more in-depth tutorial can
be accessed at:
\url{https://jianhuig.github.io/fairmetrics/articles/fairmetrics.html}.
All code is open-source and hosted on GitHub. All bugs and inquiries can
be reported at \url{https://github.com/jianhuig/fairmetrics/issues/}.

\bibliographystyle{plainnat}
\bibliography{references}

\end{document}